\documentstyle[aaspp4,12pt,epsf]{article}

\begin{document}

\title{Rate coefficients for rovibrational transitions in H$_2$ 
due to collisions with He} 
\author{N. Balakrishnan, M. Vieira\footnote{ 
Present address: Department of Physics and Astronomy, University of Rochester},
J. F. Babb, R. C. Forrey, and A. Dalgarno}
\affil{Institute for Theoretical Atomic and Molecular Physics\\
Harvard-Smithsonian Center for Astrophysics\\
60 Garden Street, Cambridge, MA 02138}
\author{S. Lepp}
\affil{Department of Physics, 4505 South Maryland Parkway\\
University of Nevada, Las Vegas, NV 89154}

\begin{abstract}
We present quantum mechanical and quasiclassical trajectory calculations 
of cross sections for rovibrational transitions 
in ortho- and para-H$_2$ induced by collisions with He atoms.
Cross sections were
obtained for kinetic energies between 10$^{-4}$ and 3 eV, and 
the corresponding rate
coefficients were calculated for the temperature range
100${\leq}T{\leq}$4000 K. 
 Comparisons are made with previous
calculations. 
\end{abstract}
\clearpage
\section{Introduction}
Rovibrationally excited H$_2$ molecules have been observed in 
many astrophysical objects (for recent studies, see Weintraub et al. 1998; 
van Dishoeck et al. 1998; Shupe et al. 1998; Bujarrabal et al. 1998; 
Stanke et al. 1998).
 The rovibrational levels
of the molecule  may be
populated by ultraviolet pumping, by X-ray pumping, by the formation
mechanism, and by collisional excitation in shock-heated gas (Dalgarno
1995).  The excited level populations are then modified by
collisions followed by quadrupole emissions. 
The main colliding partners apart from H$_2$ are H and He.
Although He is
only one tenth as abundant as H, collisions with He may have a
significant influence in many astronomical environments depending 
on the density, temperature and the 
 initial rotational and vibrational excitation of 
the molecule. Collisions with He and H$_2$ are particularly important 
when most of the hydrogen is in molecular form, as in dense molecular clouds. 
To interpret observations of the radiation
emitted by the gas, the collision cross sections and 
corresponding rate coefficients
characterizing the collisions must be known. 
Emissions from excited rovibrational levels
of the molecule provide important
clues regarding the physical state of the gas, dissociation,
excitation and formation properties of H$_2$.  Here we investigate 
the collisional relaxation of vibrationally excited H$_2$ by He.

Rovibrational transitions in H$_2$ induced by collisions 
with He atoms have been 
the subject of a large number of theoretical calculations 
in the past (Alexander 1976, 1977; Alexander and McGuire 1976; 
Dove et al. 1980;
Eastes and Secrest 1972; Krauss and Mies 1965; McGuire and Kouri 1974; 
Raczkowski et al. 1978)
and continue 
to attract experimental (Audibert et al. 1976; 
Michaut et al. 1998) and theoretical attention
(Flower et al. 1998; Dubernet \& Tuckey 1999; Balakrishnan et al. 1999).
Recent theoretical calculations 
 are motivated by  the 
availability of more accurate representations
 of the interaction potentials 
and the possibility of performing 
quantum mechanical calculations with few
approximations. The potential energy surface determined by
Muchnick and Russek (1994)  was used by Flower et al. (1998) and by 
Balakrishnan et al. (1999) in recent 
quantum mechanical  calculations of rovibrational transition 
rate coefficients for temperatures ranging from 
100 to 5000K. Flower et al. presented their results
for vibrational levels $v=0,1$ and 2  of 
ortho- and para-H$_2$.  Balakrishnan 
et al. (1999) reported similar results for $v=0$ and 1.
Though both authors have adopted  similar close-coupling 
approaches for the scattering calculations, Flower et al. 
used a harmonic oscillator approximation for H$_2$ vibrational 
wave functions in evaluating the matrix elements of the 
potential while the calculations of Balakrishnan et al. made use of 
the H$_2$ potential of Schwenke (1988) and the corresponding 
numerically determined wave functions. The results of the two calculations
agreed well for pure rotational transitions but 
some discrepancies were seen 
for rovibrational transitions. We believe this 
may be due to the different choice of  vibrational 
wave functions.
The sensitivity of 
the rate coefficients to the choice of the H$_2$ wave function 
was  noted previously and differences could be 
significant for excited vibrational levels. We find this to be 
the case for transitions involving $v\geq 2$. 
 Thus, in this article, we report 
rate coefficients for transitions from  $v=2$ to 6 initial states 
of H$_2$ induced by collisions with He atoms 
using numerically exact quantum mechanical calculations. 
 We  also report results of 
quasiclassical trajectory (QCT) calculations  and 
examine the suitability of classical mechanical calculations  in predicting 
rovibrational transitions in H$_2$.

\section{Results}

The quantum mechanical calculations were performed using the
nonreactive scattering program MOLSCAT developed by Hutson and Green
(1994) with the He-H$_2$ interaction potential of Muchnick and Russek
(1994) and the H$_2$ potential of Schwenke (1988). 
We refer to our earlier paper (Balakrishnan, Forrey \& Dalgarno, 1999) for 
details of the numerical implementation.
Different  basis sets were used in the calculations for transitions from 
different initial vibrational levels. We use the 
notation [$v_1$--$v_2$]($j_1$--$j_2$) to represent the basis set
 where the quantities within 
the square brackets give the range 
of vibrational levels and those in braces give the range of
rotational levels coupled in each of the vibrational levels.
For transitions from $v=2,3$ and 4 we used, respectively,  
the basis sets [0--3](0--11) \& [4](0--3),
[0--3](0--11) \& [4](0--9) and [3--5](0--11) \& [1,6](0--11). For 
$v=5$ and 6 of para H$_2$ we used, respectively,   
[4--6](0--14) \& [3,7](0--8) 
and [5--7](0--14) \& [4,8](0--8).
During the calculations, we found that 
the $\Delta v=\pm 2$ transitions are weak with cross sections 
that are typically 
orders of magnitude smaller 
than for the $\Delta v=\pm 1$ transitions.
Thus, for $v=5$ and 6 of ortho-H$_2$, we have only included 
the $\Delta v=\pm 1$ vibrational levels with $j$=0--13 in the basis set to 
reduce the computational effort.
The basis sets were chosen as a compromise between 
numerical efficiency and accuracy and could introduce some truncation
errors for transitions to levels which lie at the outer edge 
of the basis set. Our convergence tests show that truncation errors are small.  
Rovibrational transition cross sections 
$\sigma_{vj,v'j'}$ where the pairs of numbers $vj$ and $v'j'$  respectively denote 
the initial and final rovibrational quantum numbers, were computed for 
kinetic energies ranging from 10$^{-4}$ to 3 eV. 
Sufficient total angular momentum
partial waves were included in the calculations to secure convergence of the 
cross sections.

The quasiclassical calculations were carried out 
using the standard classical trajectory method
as described by Lepp, Buch and Dalgarno (1995) in which the 
procedure of Blais and Truhlar (1976) was adopted for the final 
state analysis. Because rovibrational transitions are rare 
at low velocities, useful results could be obtained only for collisions 
at energies above 0.1 eV. The results are averages over 10000 trajectories.
The quantum mechanical calculations were performed using MOLSCAT (Hutson \& Green, 1994)
suitably adapted for the present system with the potential
represented by a Legendre polynomial expansion in which we retained 
nonvanishing terms of orders 0 to 10, inclusive.

Calculation of rovibrational transition rate coefficients over 
a wide range of temperatures requires the determination of scattering 
cross sections at the energies spanned by the
Boltzmann distribution at each temperature. This is a computationally demanding 
problem especially when quantum mechanical calculations are required.
For many systems, QCT calculations offer a
good compromise between accuracy and computational effort. However, 
the validity of classical mechanics is in question especially for 
lighter systems and at lower temperatures where quantum mechanical effects such as 
tunneling are important. Due to the small masses of the atoms involved, 
the present system offers an excellent opportunity to test the reliability 
of QCT calculations in predicting rovibrational transitions. 
Though such attempts have been made in the past 
(Dove et al. 1980) the classical mechanical and quantum mechanical calculations 
were done in different energy regimes and a one-to-one comparison 
was not possible.

We carried out quantum mechanical
and QCT calculations of rovibrational transition cross 
sections for the present system over a wide range of energies. 
In  Figure 1 we compare the results  for
pure rotational de-excitation transitions ($\Delta j=j'-j=-2$)
with $j=2$ in $v=0$ to 5. 
There are striking similarities and 
differences between the quantum mechanical and  QCT
results. 
  They agree quite well at higher energies and have a similar
energy dependence with both 
calculations predicting the same maximum for the cross sections before they 
fall off. In contrast,  the agreement becomes less satisfactory 
at lower energies with the QCT cross sections rapidly 
decreasing to zero as the energy is decreased. The quantum mechanical 
results pass through  minima
 and subsequently increase with further decrease of the kinetic 
energy. This is a purely quantum mechanical effect 
that has  important 
consequences for the low temperature rate coefficients (Balakrishnan et al. 1998).
 The quantum mechanical 
cross sections eventually conform to an 
inverse velocity dependence as the relative translational
energy is decreased to zero and the corresponding rate coefficients are
finite in  the limit of zero temperature, in accordance with Wigner's threshold law. 
The results illustrate that QCT calculations may be 
used reliably to calculate rotational transitions for energies higher than 0.5 eV, but
at lower  energies, 
quantum mechanical calculation
must be employed. The energy regime for the validity of QCT results
is more restricted for rovibrational transitions
as illustrated in  Figure 2 in which 
we show cross sections for rovibrational transitions
$v,j\rightarrow v-1,j+2$ with $j=5$  in 
$v=3$ to 5. The results  indicate that 
the QCT method is inadequate to calculate rovibrational 
transition cross sections at impact energies below 1 eV. The QCT 
results exhibit a sharp fall below a threshold near an energy of 1 eV whereas 
the quantum mechanical results vary smoothly. 
Similar results hold for other 
transitions. The higher collision energies required for the validity 
of the QCT method for rovibrational transitions compared to 
pure rotational transitions is due in part to the much smaller  
cross sections for rovibrational transitions making the results
more sensitive to the details of the dynamics.

Rate coefficients $k_{vj,v'j'}(T)$ were calculated for the temperature range 
$T=100$ to 4000 K by averaging the cross sections over 
a Boltzmann distribution of relative velocities.
Flower et al. (1998) have reported rate coefficients for rovibrational 
transitions from $v=0-2$ for $T=1000, 2000$ and 4500 K.
A comparison of some of the de-excitation 
rate coefficients from the $v=2$ level calculated in this paper and
those reported by Flower {\it et al.} is given in Table 1 for $T=1000$ K. 
The
agreement  is good for rovibrational transitions involving $|\Delta j|=0,2$ and 4
but larger differences,  by factors between 2 and 4, are seen for
transitions involving larger values of $|\Delta j|$.
 Our rate coefficients are generally greater than those of Flower {\it et al.} (1998)
 for  transitions where the discrepancy is large.
  Our rate coefficient calculations extend those of
Flower {\it et al.} (1998) to include the $v=4,5$ and 6 vibrational levels. The 
preferential formation of H$_2$
in these vibrational levels has been discussed by Dalgarno (1995). 

In Tables 2 and 3, we present our comprehensive results for 
rovibrational de-excitation transitions from different rotational 
levels in para- and ortho-H$_2$ from the $v=2$ level as 
functions of temperature. The corresponding excitation rate coefficients 
may be obtained from detailed balance
\begin{equation}
k_{v'j',v,j}(T)=\frac{(2j+1)}{(2j'+1)} \exp{[(\epsilon_{v'j'}-
\epsilon_{vj})/k_BT]} k_{vj,v'j'}(T)
\end{equation}
where $k_B$ is the Boltzmann constant and $\epsilon_{vj}$ is the 
rovibrational energy of the molecule in the $v,j$ level.
Similar results 
for $v=3$ to 6 are presented in Tables 4-11. Tables 2-11 reveal
some interesting aspects of energy transfer.
It can be seen that for temperatures less than 1000 K, 
 rate coefficients for  rovibrational transitions 
involving $\Delta j=-4\Delta v$ 
predominate over 
other transitions where changes in both $v$ and $j$ occur.
 This is clearly seen for ($vj,v'j'$)
transitions with $\Delta v=v'-v=-1$
 for $j=3-7$ and the effect becomes stronger with increasing 
rotational excitation but less important as $T$ increases. 
This is an example of  quasi-resonant scattering
(Stewart et al. 1988; Forrey et al. 1999). A detailed 
study of this process in the limit of zero temperature 
and its
correspondence with classical mechanics has been carried out recently (Forrey et al. 1999).
The efficient conversion of vibrational energy into rotational energy 
may produce a significant population of high rotational levels
in environments where molecular hydrogen is subjected to an intense flux 
of X-rays or ultraviolet photons.

\section{Acknowledgments}
This work was supported by the
 National Science Foundation (NSF), Division of Astronomy. MV was supported 
by the NSF through the Research Experience for Undergraduates
 program at the Smithsonian Astrophysical Observatory.

\clearpage

\renewcommand{\arraystretch}{1.0}
\setlength{\tabcolsep}{.05in}
\begin{table}[h]
\caption{A comparison of rate coefficients (in units of
cm$^3$s$^{-1}$) for collisionally induced rovibrational transitions in para-H$_2$
computed in this paper and those of Flower et al. (1998) at $T=1000$ K.} 
\end{table}
\begin{center}
\begin{tabular}{ccc}
\hline\hline
${vj},{v'j'}$ & This Work & Flower \it{et al.} \\
\hline
20,10 &  1.24(-15) & 1.1(-15) \\  
20,12 &  2.03(-15) & 2.6(-15) \\  
20,14 &  3.65(-15) & 3.7(-15) \\  
20,16 &  3.28(-15) & 1.3(-15) \\  
20,18 &  7.75(-16) & 1.6(-16) \\  
22,10 &  3.64(-16) & 3.4(-16) \\ 
22,12 &  2.87(-15) & 2.8(-15) \\  
22,14 &  8.05(-15) & 7.2(-15) \\  
22,16 &  5.16(-15) & 4.2(-15) \\  
22,18 &  2.14(-15) & 6.3(-16) \\  
22,20 &  2.45(-11) & 1.8(-11) \\  
21,11 &  1.82(-15) & 1.8(-15) \\
21,13 &  4.38(-15) & 4.6(-15) \\
21,15 &  3.95(-15) & 3.4(-15) \\
21,17 &  2.52(-15) & 8.0(-16) \\
23,11 &  9.48(-16) & 8.6(-16) \\
23,13 &  4.42(-15) & 3.7(-15) \\
23,15 &  1.41(-14) & 1.2(-14) \\
23,17 &  1.19(-14) & 7.7(-15) \\
23,21 &  2.55(-11) & 1.8(-11) \\

\hline
\end{tabular}
\end{center}
\clearpage
\renewcommand{\arraystretch}{0.9}
\setlength{\tabcolsep}{.05in}
\begin{table}[h]
\caption{Rate coefficients (in units of
cm$^3$s$^{-1}$) for collisionally induced rovibrational transitions in
para-H$_2$ at different values of the translational
temperature $T$.}  
\end{table}
\begin{center}
\begin{tabular}{ccccccccc}\\
& \multicolumn{8}{c}{$T(K)$} \\ \hline\hline
${vj,v'j'}$&100&200&300&500&1000&2000&3000&4000\\ 
\hline
20,10 &2.81(-19)&2.27(-18)&9.84(-18)& 7.51(-17) & 1.24(-15) &1.67(-14)& 6.93(-14)&1.80(-13) \\  
20,12 &9.72(-19)&6.91(-18)&2.80(-17)& 1.76(-16) & 2.03(-15) &2.44(-14)& 1.05(-13)&2.80(-13) \\  
20,14 &1.05(-19)&1.36(-18)&9.47(-18)& 1.29(-16) & 3.65(-15) &5.64(-14)& 2.11(-13)&4.88(-13) \\  
20,16 &1.39(-18)&8.83(-18)&3.23(-17)& 1.94(-16) & 3.28(-15) &6.83(-14)& 3.10(-13)&7.69(-13) \\  
20,18 &1.20(-19)&9.40(-19)&2.56(-18)& 1.96(-17) & 7.75(-16) &2.92(-14)& 2.03(-13)&6.54(-13) \\  
22,10 &6.41(-20)&2.14(-19)&8.50(-19)& 1.27(-17) & 3.64(-16) &6.39(-15)& 2.67(-14)&6.62(-14) \\ 
22,12 &6.45(-19)&2.72(-18)&1.09(-17)& 1.26(-16) & 2.87(-15) &4.25(-14)& 1.70(-13)&4.19(-13) \\  
22,14 &2.89(-18)&1.80(-17)&7.27(-17)& 5.46(-16) & 8.05(-15) &8.79(-14)& 3.06(-13)&6.87(-13) \\  
22,16 &3.88(-18)&2.36(-17)&8.03(-17)& 3.99(-16) & 5.16(-15) &8.97(-14)& 3.82(-13)&9.27(-13) \\  
22,18 &4.49(-17)&8.25(-17)&1.17(-16)& 2.38(-16) & 2.14(-15) &4.58(-14)& 2.71(-13)&8.12(-13) \\  
22,20 &1.68(-12)&3.71(-12)&5.98(-12)& 1.10(-11) & 2.45(-11) &4.69(-11)& 6.24(-11)&7.36(-11) \\  
24,10 &9.46(-21)&7.68(-20)&4.14(-19)& 4.68(-18) & 1.60(-16) &3.91(-15)& 1.87(-14)&4.86(-14) \\  
24,12 &1.14(-19)&9.15(-19)&4.70(-18)& 4.86(-17) & 1.41(-15) &2.82(-14)& 1.22(-13)&2.99(-13) \\  
24,14 &8.87(-19)&6.94(-18)&3.26(-17)& 2.88(-16) & 6.16(-15) &9.36(-14)& 3.61(-13)&8.44(-13) \\  
24,16 &1.02(-17)&7.03(-17)&2.81(-16)& 1.90(-15) & 2.53(-14) &2.52(-13)& 8.10(-13)&1.72(-12) \\ 
24,18 &2.63(-16)&1.08(-15)&2.71(-15)& 8.61(-15) & 3.62(-14) &1.90(-13)& 6.59(-13)&1.55(-12) \\  
24,20 &9.46(-15)&3.58(-14)&8.74(-14)& 2.83(-13) & 1.32(-12) &4.83(-12)& 8.72(-12)&1.22(-11) \\  
24,22 &2.98(-13)&9.52(-13)&2.02(-12)& 5.31(-12) & 1.82(-11) &4.96(-11)& 7.85(-11)&1.03(-10) \\  
26,10 &1.11(-21)&1.15(-20)&6.67(-20)& 8.71(-19) & 4.39(-17) &1.69(-15)& 1.03(-14)&3.09(-14) \\  
26,12 &1.26(-20)&1.27(-19)&7.17(-19)& 8.90(-18) & 3.93(-16) &1.29(-14)& 7.14(-14)&2.00(-13) \\  
26,14 &1.13(-19)&1.04(-18)&5.57(-18)& 6.14(-17) & 2.03(-15) &4.86(-14)& 2.27(-13)&5.77(-13) \\  
26,16 &1.60(-18)&1.29(-17)&6.08(-17)& 5.44(-16) & 1.18(-14) &1.84(-13)& 7.02(-13)&1.60(-12) \\  
26,18 &5.89(-17)&3.70(-16)&1.37(-15)& 8.29(-15) & 9.59(-14) &8.22(-13)& 2.35(-12)&4.50(-12) \\  
26,110&1.31(-14)&3.07(-14)&5.17(-14)& 9.89(-14) & 2.22(-13) &5.31(-13)& 1.29(-12)&2.72(-12) \\
26,20 &6.77(-17)&3.80(-16)&1.28(-15)& 6.85(-15) & 6.87(-14) &5.18(-13)& 1.34(-12)&2.31(-12) \\  
26,22 &1.42(-15)&7.24(-15)&2.21(-14)& 1.02(-13) & 8.00(-13) &4.63(-12)& 1.05(-11)&1.68(-11) \\  
26,24 &4.64(-14)&1.86(-13)&4.60(-13)& 1.55(-12) & 7.81(-12) &2.96(-11)& 5.41(-11)&7.68(-11) \\  
\hline
\end{tabular}
\end{center}
\clearpage
\begin{table}[h]
\caption{Same as Table 2 except for ortho-H$_2$.}
\end{table}
\begin{center}
\begin{tabular}{ccccccccc}\\ 
& \multicolumn{8}{c}{$T(K)$} \\ \hline\hline
${vj,v'j'}$&100&200&300&500&1000&2000&3000&4000\\ 
\hline
21,11 &2.45(-19) &2.25(-18)&1.04(-17)& 9.14(-17) & 1.82(-15)&2.75(-14) & 1.16(-13)&2.97(-13) \\
21,13 &1.42(-18) &1.09(-17)&4.66(-17)& 3.28(-16) & 4.38(-15)&4.69(-14) & 1.72(-13)&4.12(-13) \\
21,15 &6.32(-19)&4.64(-18)&1.60(-17)& 1.32(-16) & 3.95(-15)&7.51(-14) & 3.01(-13)&7.02(-13) \\
21,17 &8.43(-19)&5.24(-18)&2.23(-17)& 1.58(-16) & 2.52(-15)&5.66(-14) & 2.93(-13)&7.85(-13) \\
21,19 &6.90(-19)&1.67(-18)&2.07(-18)& 6.96(-18) & 3.28(-16)&1.52(-14) & 1.30(-13)&4.73(-13) \\
23,11 &6.70(-20)&1.04(-18)&5.29(-18)& 4.16(-17) & 9.48(-16)&1.74(-14) & 7.33(-14)&1.80(-13) \\
23,13 & 5.18(-19)&7.20(-18)&3.47(-17)& 2.46(-16) & 4.42(-15)&6.56(-14) & 2.54(-13)&6.01(-13) \\
23,15 &4.76(-18)&2.24(-17)&1.05(-16)& 9.22(-16) & 1.41(-14)&1.49(-13) & 5.02(-13)&1.09(-12) \\
23,17 &3.15(-17)&1.80(-16)&5.32(-16)& 1.98(-15) & 1.19(-14)&1.18(-13) & 4.73(-13)&1.14(-12) \\
23,19 &1.74(-17)&2.64(-17)&2.44(-17)& 7.30(-17) & 1.81(-15)&3.76(-14) & 2.34(-13)&7.32(-13) \\
23,21 &7.73(-13)&2.12(-12)&4.03(-12)& 9.09(-12) & 2.55(-11)&5.93(-11) & 8.67(-11)&1.08(-10) \\
25,11 &3.36(-21)&1.25(-19)&6.81(-19)& 8.38(-18) & 3.34(-16)&9.44(-15) & 4.83(-14)&1.30(-13) \\
25,13 &3.68(-20)&9.71(-19)&4.92(-18)& 5.35(-17) & 1.75(-15)&3.86(-14) & 1.72(-13)&4.26(-13) \\
25,15 &7.05(-19)&8.96(-18)&4.23(-17)& 3.76(-16) & 8.51(-15)&1.31(-13) & 4.98(-13)&1.14(-12) \\
25,17 &1.97(-17)&1.42(-16)&5.78(-16)& 3.80(-15) & 4.82(-14)&4.46(-13) & 1.34(-12)&2.69(-12) \\
25,19 &1.74(-15)&5.74(-15)&1.27(-14)& 3.38(-14) & 1.04(-13)&3.28(-13) & 8.92(-13)&1.93(-12) \\
25,21 &2.81(-15)&1.46(-14)&4.69(-14)& 2.02(-13) & 1.26(-12)&5.89(-12) & 1.21(-11)&1.82(-11) \\
25,23 &7.59(-14)&3.04(-13)&8.30(-13)& 2.83(-12) & 1.21(-11)&3.86(-11) & 6.57(-11)&8.97(-11) \\
27,11 &7.46(-22)&1.06(-20)&7.90(-20)& 1.30(-18) & 7.34(-17)&3.33(-15) & 2.23(-14)&7.02(-14) \\
27,13 &6.52(-21)&8.65(-20)&6.02(-19)& 8.89(-18) & 4.15(-16)&1.49(-14) & 8.65(-14)&2.48(-13) \\
27,15 &7.60(-20)&9.01(-19)&5.62(-18)& 6.93(-17) & 2.34(-15)&5.81(-14) & 2.76(-13)&7.02(-13) \\
27,17 &1.22(-18)&1.71(-17)&8.62(-17)& 7.31(-16) & 1.59(-14)&2.49(-13) & 9.32(-13)&2.07(-12) \\
27,19 &1.28(-16)&8.18(-16)&3.14(-15)& 1.87(-14) & 1.93(-13)&1.47(-12) & 3.91(-12)&7.10(-12) \\
27,111&1.41(-14)&2.79(-14)&4.96(-14)& 1.10(-13) & 2.76(-13)&6.45(-13) & 1.51(-12)&3.12(-12) \\
27,21 &1.52(-17)&1.35(-16)&6.40(-16)& 4.63(-15) & 6.16(-14)&6.02(-13) & 1.77(-12)&3.33(-12) \\
27,23 &2.32(-16)&1.87(-15)&8.06(-15)& 4.96(-14) & 4.84(-13)&3.35(-12) & 8.23(-12)&1.39(-11) \\
27,25 &9.84(-15)&5.68(-14)&1.95(-13)& 8.65(-13) & 5.19(-12)&2.24(-11) & 4.39(-11)&6.47(-11) \\
\hline
\end{tabular}
\end{center}
\clearpage
\begin{table}[h]
\caption{Same as Table 2 but for the $v=3$ level of para-H$_2$.}
\end{table}
\begin{center}
\begin{tabular}{ccccccccc}\\ 
& \multicolumn{8}{c}{$T(K)$} \\ \hline\hline
${vj,v'j'}$&100&200&300&500&1000&2000&3000&4000\\ 
\hline
30,20 &1.07(-18)&8.34(-18)&3.53(-17)& 2.47(-16) & 3.49(-15)&4.21(-14) & 1.66(-13)&4.08(-13) \\
30,22 &3.26(-18)&2.24(-17)&8.59(-17)& 4.87(-16) & 4.90(-15)&5.66(-14) & 2.34(-13)&5.94(-13) \\
30,24 &3.10(-19)&5.35(-18)&2.79(-17)& 3.64(-16) & 9.40(-15)&1.21(-13) & 4.20(-13)&9.36(-13) \\
30,26 &4.37(-18)&2.94(-17)&1.09(-16)& 6.47(-16) & 1.01(-14)&1.68(-13) & 6.58(-13)&1.50(-12) \\
30,28 &4.98(-19)&3.76(-18)&1.17(-17)& 9.42(-17) & 3.05(-15)&9.39(-14) & 5.37(-13)&1.50(-12) \\   
32,20 &2.40(-19)&1.04(-18)&4.18(-18)& 5.00(-17) & 1.17(-15)&1.64(-14) & 6.15(-14)&1.43(-13) \\
32,22 &2.00(-18)&1.08(-17)&4.43(-17)& 4.24(-16) & 7.98(-15)&1.01(-13) & 3.78(-13)&8.92(-13) \\
32,24 &8.28(-18)&5.81(-17)&2.37(-16)& 1.59(-15) & 1.97(-14)&1.88(-13) & 6.11(-13)&1.31(-12) \\
32,26 &1.71(-17)&9.45(-17)&3.00(-16)& 1.41(-15) & 1.59(-14)&2.19(-13) & 8.18(-13)&1.83(-12) \\   
32,28 &1.51(-16)&2.68(-16)&3.84(-16)& 8.19(-16) & 7.08(-15)&1.33(-13) & 6.72(-13)&1.78(-12) \\   
32,30 &2.10(-12)&4.67(-12)&7.63(-12)& 1.41(-11) & 3.00(-11)&5.35(-11) & 6.87(-11)&7.93(-11) \\   
34,20 &4.51(-20)&3.71(-19)&1.96(-18)& 2.07(-17) & 5.97(-16)&1.13(-14) & 4.61(-14)&1.08(-13) \\
34,22 &4.81(-19)&3.94(-18)&1.98(-17)& 1.91(-16) & 4.66(-15)&7.41(-14) & 2.79(-13)&6.32(-13) \\
34,24 &3.21(-18)&2.52(-17)&1.15(-16)& 9.63(-16) & 1.77(-14)&2.24(-13) & 7.83(-13)&1.72(-12) \\
34,26 &3.09(-17)&2.08(-16)&8.09(-16)& 5.18(-15) & 6.15(-14)&5.39(-13) & 1.62(-12)&3.27(-12) \\   
34,28 &8.59(-16)&3.38(-15)&8.20(-15)& 2.46(-14) & 9.56(-14)&4.64(-13) & 1.45(-12)&3.12(-12) \\   
34,30 &1.69(-14)&6.25(-14)&1.49(-13)& 4.67(-13) & 2.02(-12)&6.64(-12) & 1.11(-11)&1.48(-11) \\   
34,32 &4.41(-13)&1.38(-12)&2.86(-12)& 7.29(-12) & 2.37(-11)&6.05(-11) & 9.19(-11)&1.17(-10) \\   
36,20 &5.40(-21)&6.41(-20)&3.79(-19)& 4.77(-18) & 1.99(-16)&5.90(-15) & 2.96(-14)&7.69(-14) \\
36,22 &6.08(-20)&6.51(-19)&3.73(-18)& 4.44(-17) & 1.62(-15)&4.10(-14) & 1.88(-13)&4.65(-13) \\
36,24 &5.01(-19)&4.60(-18)&2.43(-17)& 2.55(-16) & 7.12(-15)&1.35(-13) & 5.40(-13)&1.23(-12) \\
36,26 &5.97(-18)&4.67(-17)&2.15(-16)& 1.82(-15) & 3.42(-14)&4.42(-13) & 1.51(-12)&3.19(-12) \\   
36,28 &1.71(-16)&1.06(-15)&3.81(-15)& 2.18(-14) & 2.26(-13)&1.69(-12) & 4.48(-12)&8.16(-12) \\  
36,210&3.35(-14)&7.60(-14)&1.26(-13)& 2.36(-13) & 5.03(-13)&1.20(-12) & 2.85(-12)&5.56(-12) \\
36,30 &1.65(-16)&9.11(-16)&3.02(-15)& 1.54(-14) & 1.38(-13)&8.91(-13) & 2.05(-12)&3.27(-12) \\   
36,32 &2.88(-15)&1.46(-14)&4.43(-14)& 1.94(-13) & 1.37(-12)&6.97(-12) & 1.45(-11)&2.19(-11) \\   
36,34 &7.49(-14)&3.00(-13)&7.43(-13)& 2.41(-12) & 1.09(-11)&3.75(-11) & 6.50(-11)&8.92(-11) \\   
\hline
\end{tabular}
\end{center}
\clearpage
\begin{table}[h]
\caption{Same as Table 4 but for ortho-H$_2$.}
\end{table}
\begin{center}
\begin{tabular}{ccccccccc}\\ 
& \multicolumn{8}{c}{$T(K)$} \\ \hline\hline
${vj,v'j'}$&100&200&300&500&1000&2000&3000&4000\\ 
\hline
31,21 &9.61(-19)&8.24(-18)&3.76(-17)& 3.05(-16) & 5.20(-15)&6.88(-14) & 2.72(-13)&6.68(-13) \\
31,23 &4.95(-18)&3.52(-17)&1.45(-16)& 9.36(-16) & 1.06(-14)&1.01(-13) & 3.56(-13)&8.23(-13) \\
31,25 &3.48(-18)&2.22(-17)&7.27(-17)& 5.03(-16) & 1.15(-14)&1.72(-13) & 6.12(-13)&1.33(-12) \\
31,27 &3.94(-18)&1.97(-17)&7.89(-17)& 5.43(-16) & 8.16(-15)&1.51(-13) & 6.61(-13)&1.58(-12) \\
31,29 &4.17(-18)&9.92(-18)&1.23(-17)& 3.66(-17) & 1.44(-15)&5.32(-14) & 3.68(-13)&1.14(-12) \\ 
33,21 &3.64(-19)&3.26(-18)&1.63(-17)& 1.45(-16) & 3.04(-15)&4.50(-14) & 1.68(-13)&3.82(-13) \\
33,23 &2.04(-18)&1.17(-17)&5.84(-17)& 6.25(-16) & 1.23(-14)&1.54(-13) & 5.46(-13)&1.23(-12) \\
33,25 &1.50(-17)&9.82(-17)&3.93(-16)& 2.69(-15) & 3.41(-14)&3.19(-13) & 9.97(-13)&2.05(-12) \\
33,27 &1.14(-16)&5.67(-16)&1.63(-15)& 5.99(-15) & 3.42(-14)&2.91(-13) & 1.03(-12)&2.26(-12) \\
33,29 &7.14(-18)&5.74(-17)&1.18(-16)& 4.21(-16) & 6.07(-15)&1.10(-13) & 5.85(-13)&1.61(-12) \\ 
33,31 &1.10(-12)&2.85(-12)&5.21(-12)& 1.15(-11) & 3.16(-11)&6.99(-11) & 9.85(-11)&1.20(-10) \\
35,21 &6.90(-20)&6.45(-19)&3.58(-18)& 4.15(-17) & 1.32(-15)&2.84(-14) & 1.22(-13)&2.91(-13) \\
35,23 &4.97(-19)&4.37(-18)&2.27(-17)& 2.34(-16) & 6.01(-15)&1.03(-13) & 3.96(-13)&8.88(-13) \\
35,25 &4.36(-18)&3.42(-17)&1.58(-16)& 1.34(-15) & 2.45(-14)&3.11(-13) & 1.06(-12)&2.27(-12) \\
35,27 &6.99(-17)&4.47(-16)&1.69(-15)& 1.03(-14) & 1.15(-13)&9.30(-13) & 2.61(-12)&4.96(-12) \\
35,29 &6.11(-15)&1.88(-14)&3.81(-14)& 9.04(-14) & 2.52(-13)&7.54(-13) & 1.89(-12)&3.77(-12) \\ 
35,31 &8.21(-15)&3.58(-14)&9.78(-14)& 3.70(-13) & 2.07(-12)&8.61(-12) & 1.63(-11)&2.33(-11) \\
35,33 &1.81(-13)&6.43(-13)&1.48(-12)& 4.31(-12) & 1.65(-11)&4.83(-11) & 7.85(-11)&1.04(-10) \\ 
37,21 &7.60(-21)&8.23(-20)&5.12(-19)& 7.35(-18) & 3.58(-16)&1.23(-14) & 6.65(-14)&1.79(-13) \\
37,23 &5.66(-20)&5.81(-19)&3.41(-18)& 4.40(-17) & 1.76(-15)&4.82(-14) & 2.30(-13)&5.67(-13) \\
37,25 &5.33(-19)&4.96(-18)&2.63(-17)& 2.83(-16) & 8.20(-15)&1.60(-13) & 6.45(-13)&1.47(-12) \\
37,27 &8.72(-18)&6.89(-17)&3.12(-16)& 2.54(-15) & 4.66(-14)&5.88(-13) & 1.95(-12)&4.00(-12) \\
37,29 &4.85(-16)&2.81(-15)&9.55(-15)& 4.92(-14) & 4.45(-13)&2.96(-12) & 7.25(-12)&1.24(-11) \\
37,211&4.72(-14)&1.02(-13)&1.64(-13)&2.95(-13) & 6.10(-13)&1.41(-12) & 3.24(-12)&6.21(-12) \\
37,31 &7.94(-17)&5.07(-16)&1.90(-15)& 1.15(-14) & 1.33(-13)&1.11(-12) & 2.90(-12)&4.99(-12) \\
37,33 &1.04(-15)&5.91(-15)&1.99(-14)& 9.99(-14) & 8.52(-13)&5.17(-12) & 1.16(-11)&1.84(-11) \\
37,35 &3.37(-14)&1.47(-13)&3.95(-13)& 1.43(-12) & 7.48(-12)&2.91(-11) & 5.35(-11)&7.58(-11) \\
\hline
\end{tabular}
\end{center}
\clearpage
\begin{table}[h]
\caption{Same as Table 2 but for the $v=4$ level of para-H$_2$.}
\end{table} 
\begin{center}
\begin{tabular}{ccccccccc} \\ 
& \multicolumn{8}{c}{$T(K)$} \\ \hline\hline
${vj,v'j'}$&100&200&300&500&1000&2000&3000&4000\\ 
\hline
40,30 &3.80(-18)&2.84(-17)&1.15(-16)& 7.34(-16) & 8.72(-15)&9.41(-14) & 3.45(-13)&7.93(-13)\\
40,32 &1.04(-17)&6.85(-17)&2.42(-16)& 1.22(-15) & 1.09(-14)&1.21(-13) & 4.68(-13)&1.13(-12)\\
40,34 &3.10(-18)&2.65(-17)&1.27(-16)& 1.24(-15) & 2.35(-14)&2.44(-13) & 7.93(-13)&1.73(-12)\\
40,36 &1.18(-17)&8.57(-17)&3.20(-16)& 1.89(-15) & 2.76(-14)&3.72(-13) & 1.31(-12)&2.85(-12)\\
40,38 &2.13(-18)&1.25(-17)&4.47(-17)& 3.58(-16) & 9.81(-15)&2.64(-13) & 1.31(-12)&3.27(-12)\\
42,30 &8.21(-19)&4.63(-18)&1.92(-17)& 1.86(-16) & 3.42(-15)&3.82(-14) & 1.28(-13)&2.78(-13)\\
42,32 &5.78(-18)&3.93(-17)&1.68(-16)& 1.35(-15) & 2.06(-14)&2.21(-13) & 7.65(-13)&1.70(-12)\\
42,34 &2.28(-17)&1.73(-16)&7.04(-16)& 4.29(-15) & 4.52(-14)&3.77(-13) & 1.16(-12)&2.42(-12)\\
42,36 &6.41(-17)&3.31(-16)&9.85(-16)& 4.34(-15) & 4.36(-14)&4.83(-13) & 1.63(-12)&3.46(-12)\\
42,38 &4.36(-16)&7.42(-16)&1.08(-15)& 2.45(-15) & 2.12(-14)&3.63(-13) & 1.61(-12)&3.86(-12)\\
42,40 &2.72(-12)&5.95(-12)&9.66(-12)& 1.75(-11) & 3.54(-11)&5.96(-11) & 7.43(-11)&8.42(-11)\\
44,30 &2.00(-19)&1.68(-18)&8.65(-18)& 8.58(-17) & 2.05(-15)&3.04(-14) & 1.07(-13)&2.30(-13)\\
44,32 &2.09(-18)&1.45(-17)&7.15(-17)& 6.85(-16) & 1.41(-14)&1.80(-13) & 6.08(-13)&1.29(-12)\\
44,34 &1.24(-17)&7.53(-17)&3.37(-16)& 2.86(-15) & 4.60(-14)&4.88(-13) & 1.57(-12)&3.29(-12)\\
44,36 &9.22(-17)&5.58(-16)&2.09(-15)& 1.29(-14) & 1.36(-13)&1.05(-12) & 2.98(-12)&5.78(-12)\\
44,38 &2.44(-15)&9.32(-15)&2.19(-14)& 6.21(-14) & 2.25(-13)&1.07(-12) & 3.12(-12)&6.30(-12)\\
44,40 &2.93(-14)&1.07(-13)&2.52(-13)& 7.51(-13) & 2.96(-12)&8.63(-12) & 1.35(-11)&1.73(-11)\\
44,42 &6.42(-13)&1.99(-12)&4.07(-12)& 1.00(-11) & 3.03(-11)&7.21(-11) & 1.05(-10)&1.30(-10)\\
46,30 &3.23(-20)&3.46(-19)&2.05(-18)& 2.42(-17) & 8.33(-16)&1.90(-14) & 8.05(-14)&1.86(-13)\\
46,32 &3.15(-19)&3.23(-18)&1.85(-17)& 2.04(-16) & 6.14(-15)&1.20(-13) & 4.73(-13)&1.05(-12)\\
46,34 &2.08(-18)&1.97(-17)&1.04(-16)& 9.95(-16) & 2.30(-14)&3.49(-13) & 1.22(-12)&2.54(-12)\\
46,36 &1.79(-17)&1.13(-16)&4.95(-16)& 4.86(-15) & 9.20(-14)&1.02(-12) & 3.14(-12)&6.18(-12)\\
46,38 &4.36(-16)&2.68(-15)&9.49(-15)& 5.13(-14) & 4.81(-13)&3.24(-12) & 8.10(-12)&1.41(-11)\\
46,310&7.36(-14)&1.59(-13)&2.53(-13)& 4.28(-13) & 6.92(-13)&1.17(-12) & 2.98(-12)&6.32(-12)\\
46,40 &3.83(-16)&2.10(-15)&6.85(-15)& 3.26(-14) & 2.54(-13)&1.37(-12) & 2.81(-12)&4.16(-12)\\
46,42 &5.72(-15)&2.88(-14)&8.65(-14)& 3.58(-13) & 2.22(-12)&9.79(-12) & 1.87(-11)&2.67(-11)\\
46,44 &1.19(-13)&4.67(-13)&1.15(-12)& 3.59(-12) & 1.49(-11)&4.63(-11) & 7.61(-11)&1.01(-10)\\
\hline
\end{tabular}
\end{center}
\clearpage

\begin{table}[h]
\caption{Same as Table 6 except for ortho-H$_2$.}
\end{table}
\begin{center}
\begin{tabular}{ccccccccc} \\ 
& \multicolumn{8}{c}{$T(K)$} \\ \hline\hline
${vj,v'j'}$&100&200&300&500&1000&2000&3000&4000\\ 
\hline
41,31 &3.30(-18)&2.80(-17)&1.27(-16)& 9.43(-16) & 1.35(-14)&1.54(-13) & 5.66(-13)&1.30(-12)\\
41,33 &1.40(-17)&1.03(-16)&4.14(-16)& 2.45(-15) & 2.37(-14)&2.04(-13) & 6.77(-13)&1.49(-12)\\
41,35 &1.43(-17)&8.40(-17)&2.80(-16)& 1.69(-15) & 2.86(-14)&3.47(-13) & 1.13(-12)&2.36(-12)\\
41,37 &1.52(-17)&6.71(-17)&2.50(-16)& 1.66(-15) & 2.33(-14)&3.56(-13) & 1.37(-12)&3.07(-12)\\
41,39 &2.30(-17)&4.35(-17)&4.87(-17)& 1.36(-16) & 5.14(-15)&1.67(-13) & 9.87(-13)&2.71(-12)\\
43,31 &1.48(-18)&1.40(-17)&6.88(-17)& 5.38(-16) & 9.11(-15)&1.08(-13) & 3.55(-13)&7.47(-13)\\
43,33 &8.07(-18)&7.16(-17)&3.30(-16)& 2.31(-15) & 3.22(-14)&3.36(-13) & 1.09(-12)&2.31(-12)\\
43,35 &4.51(-17)&3.45(-16)&1.40(-15)& 7.92(-15) & 7.79(-14)&6.37(-13) & 1.86(-12)&3.68(-12)\\
43,37 &3.67(-16)&1.86(-15)&5.39(-15)& 1.88(-14) & 9.16(-14)&6.44(-13) & 2.04(-12)&4.24(-12)\\
43,39 &1.93(-16)&2.81(-16)&3.58(-16)& 1.24(-15) & 1.85(-14)&3.09(-13) & 1.45(-12)&3.62(-12)\\
43,41 &1.53(-12)&4.03(-12)&7.37(-12)& 1.57(-11) & 4.00(-11)&8.15(-11) & 1.10(-10)&1.31(-10)\\
45,31 &9.55(-19)&6.74(-18)&2.19(-17)& 1.67(-16) & 4.62(-15)&7.89(-14) & 2.92(-13)&6.32(-13)\\
45,33 &3.46(-18)&2.70(-17)&1.02(-16)& 8.09(-16) & 1.82(-14)&2.53(-13) & 8.60(-13)&1.80(-12)\\
45,35 &1.58(-17)&1.28(-16)&5.27(-16)& 3.85(-15) & 6.32(-14)&6.73(-13) & 2.11(-12)&4.26(-12)\\
45,37 &2.09(-16)&1.41(-15)&4.78(-15)& 2.47(-14) & 2.46(-13)&1.76(-12) & 4.61(-12)&8.44(-12)\\
45,39 &1.54(-14)&4.50(-14)&8.86(-14)& 2.02(-13) & 5.39(-13)&1.64(-12) & 4.03(-12)&7.63(-12)\\
45,41 &1.59(-14)&6.81(-14)&1.78(-13)& 6.29(-13) & 3.20(-12)&1.18(-11) & 2.06(-11)&2.80(-11)\\
45,43 &2.78(-13)&9.75(-13)&2.18(-12)& 6.01(-12) & 2.14(-11)&5.84(-11) & 9.08(-11)&1.17(-10)\\
47,31 &5.00(-19)&3.63(-18)&1.02(-17)& 4.71(-17) & 1.49(-15)&4.04(-14) & 1.84(-13)&4.37(-13)\\
47,33 &3.29(-19)&3.93(-18)&2.02(-17)& 1.93(-16) & 6.56(-15)&1.42(-13) & 5.74(-13)&1.29(-12)\\
47,35 &2.61(-18)&2.86(-17)&1.35(-16)& 1.07(-15) & 2.61(-14)&4.17(-13) & 1.45(-12)&2.98(-12)\\
47,37 &3.43(-17)&3.25(-16)&1.34(-15)& 8.15(-15) & 1.23(-13)&1.36(-12) & 4.06(-12)&7.72(-12)\\
47,39 &1.44(-15)&1.03(-14)&3.35(-14)& 1.31(-13) & 8.90(-13)&5.48(-12) & 1.28(-11)&2.12(-11)\\
47,311&7.76(-14)&2.02(-13)&3.36(-13)& 5.96(-13) & 1.13(-12)&2.44(-12) & 5.38(-12)&9.96(-12)\\
47,41 &1.19(-16)&7.62(-16)&3.45(-15)& 2.38(-14) & 2.59(-13)&1.77(-12) & 4.07(-12)&6.42(-12)\\
47,43 &2.24(-15)&1.23(-14)&3.94(-14)& 1.86(-13) & 1.42(-12)&7.35(-12) & 1.50(-11)&2.23(-11)\\
47,45 &6.18(-14)&2.70(-13)&6.86(-13)& 2.25(-12) & 1.04(-11)&3.64(-11) & 6.29(-11)&8.55(-11)\\
\hline
\end{tabular}
\end{center}
\clearpage

\begin{table}[h]
\caption{Same as Table 2 but for the $v=5$ level of  para-H$_2$.}
\end{table}
\begin{center}
\begin{tabular}{ccccccccc} \\
& \multicolumn{8}{c}{$T(K)$} \\ \hline\hline
${vj,v'j'}$&100&200&300&500&1000&2000&3000&4000\\ 
\hline
50,40 &1.23(-17)&8.75(-17)&3.31(-16)&1.89(-15)&1.99(-14)&1.96(-13)&6.65(-13)&1.42(-12)\\
50,42 &3.17(-17)&1.93(-16)&6.23(-16)&2.74(-15)&2.29(-14)&2.42(-13)&8.78(-13)&1.98(-12)\\
50,44 &1.64(-17)&1.12(-16)&4.80(-16)&3.73(-15)&5.15(-14)&4.44(-13)&1.34(-12)&2.77(-12)\\
50,46 &3.24(-17)&2.40(-16)&9.18(-16)&5.33(-15)&6.73(-14)&7.11(-13)&2.18(-12)&4.29(-12)\\
50,48 &6.41(-18)&3.94(-17)&1.27(-16)&1.15(-15)&3.02(-14)&5.90(-13)&2.31(-12)&4.98(-12)\\
52,40 &2.26(-18)&1.83(-17)&8.29(-17)&6.33(-16)&8.65(-15)&7.84(-14)&2.39(-13)&4.87(-13)\\
52,42 &1.64(-17)&1.26(-16)&5.42(-16)&3.80(-15)&4.74(-14)&4.42(-13)&1.41(-12)&2.94(-12)\\
52,44 &6.94(-17)&4.57(-16)&1.71(-15)&9.79(-15)&9.28(-14)&6.81(-13)&1.94(-12)&3.81(-12)\\
52,46 &2.19(-16)&1.08(-15)&3.09(-15)&1.28(-14)&1.08(-13)&9.34(-13)&2.73(-12)&5.24(-12)\\
52,48 &1.17(-15)&1.87(-15)&2.35(-15)&5.07(-15)&5.30(-14)&7.47(-13)&2.73(-12)&5.74(-12)\\
52,50 &3.56(-12)&7.67(-12)&1.22(-11)&2.14(-11)&4.10(-11)&6.52(-11)&7.91(-11)&8.83(-11)\\
54,40 &7.79(-19)&6.74(-18)&3.33(-17)&3.07(-16)&5.92(-15)&6.67(-14)&2.02(-13)&3.92(-13)\\
54,42 &6.78(-18)&5.55(-17)&2.59(-16)&2.21(-15)&3.66(-14)&3.71(-13)&1.10(-12)&2.13(-12)\\
54,44 &3.45(-17)&2.58(-16)&1.10(-15)&8.04(-15)&1.06(-13)&9.36(-13)&2.71(-12)&5.25(-12)\\
54,46 &2.34(-16)&1.50(-15)&5.47(-15)&3.04(-14)&2.76(-13)&1.84(-12)&4.74(-12)&8.54(-12)\\
54,48 &6.95(-15)&2.46(-14)&5.43(-14)&1.43(-13)&4.78(-13)&1.99(-12)&5.05(-12)&9.23(-12)\\
54,50 &5.00(-14)&1.76(-13)&4.04(-13)&1.16(-12)&4.17(-12)&1.08(-11)&1.59(-11)&1.95(-11)\\
54,52 &9.19(-13)&2.73(-12)&5.56(-12)&1.34(-11)&3.82(-11)&8.44(-11)&1.18(-10)&1.42(-10)\\
56,40 &1.54(-19)&1.45(-18)&7.81(-18)&9.33(-17)&2.81(-15)&4.66(-14)&1.60(-13)&3.21(-13)\\
56,42 &1.38(-18)&1.25(-17)&6.44(-17)&7.14(-16)&1.89(-14)&2.73(-13)&8.89(-13)&1.74(-12)\\
56,44 &7.87(-18)&6.56(-17)&3.10(-16)&2.95(-15)&6.15(-14)&7.07(-13)&2.10(-12)&3.95(-12)\\
56,46 &6.19(-17)&4.54(-16)&1.88(-15)&1.41(-14)&2.11(-13)&1.86(-12)&5.06(-12)&9.16(-12)\\
56,48 &1.10(-15)&6.32(-15)&2.06(-14)&1.03(-13)&9.03(-13)&5.29(-12)&1.21(-11)&2.00(-11)\\
56,410&1.52(-13)&3.13(-13)&5.00(-13)&8.93(-13)&1.74(-12)&3.69(-12)&7.34(-12)&1.24(-11)\\
56,50 &8.52(-16)&4.58(-15)&1.48(-14)&6.69(-14)&4.50(-13)&2.01(-12)&3.72(-12)&5.15(-12)\\
56,52 &1.09(-14)&5.33(-14)&1.59(-13)&6.33(-13)&3.50(-12)&1.34(-11)&2.35(-11)&3.18(-11)\\
56,54 &1.83(-13)&6.98(-13)&1.72(-12)&5.24(-12)&2.00(-11)&5.64(-11)&8.79(-11)&1.13(-10)\\
\hline
\end{tabular}
\end{center}

\clearpage

\begin{table}[h]
\caption{Same as Table 8 except for ortho-H$_2$.}
\end{table}
\begin{center}
\begin{tabular}{ccccccccc}\\
& \multicolumn{8}{c}{$T(K)$} \\ \hline\hline
${vj,v'j'}$&100&200&300&500&1000&2000&3000&4000\\ 
\hline
51,41 &6.91(-18)&5.48(-17)&2.48(-16)&1.86(-15)&2.71(-14)&3.17(-13)&1.15(-12)&2.60(-12)\\
51,43 &3.01(-17)&2.04(-16)&7.83(-16)&4.50(-15)&4.39(-14)&3.84(-13)&1.29(-12)&2.83(-12)\\
51,45 &3.41(-17)&1.79(-16)&6.06(-16)&3.63(-15)&5.32(-14)&5.98(-13)&1.90(-12)&3.85(-12)\\
51,47 &7.16(-17)&2.16(-16)&6.51(-16)&3.60(-15)&4.13(-14)&5.31(-13)&1.94(-12)&4.15(-12)\\
53,41 &2.75(-18)&2.17(-17)&1.02(-16)&8.29(-16)&1.39(-14)&1.69(-13)&5.85(-13)&1.27(-12)\\
53,43 &1.52(-17)&1.14(-16)&5.01(-16)&3.72(-15)&5.33(-14)&5.62(-13)&1.87(-12)&3.98(-12)\\
53,45 &8.49(-17)&5.52(-16)&2.14(-15)&1.29(-14)&1.33(-13)&1.04(-12)&2.94(-12)&5.65(-12)\\
53,47 &8.40(-16)&3.52(-15)&9.33(-15)&3.12(-14)&1.50(-13)&9.87(-13)&2.98(-12)&5.91(-12)\\
53,49 &1.15(-17)&1.32(-16)&4.77(-16)&2.99(-15)&3.34(-14)&4.09(-13)&1.69(-12)&3.95(-12)\\
53,51 &1.80(-12)&4.22(-12)&8.04(-12)&1.87(-11)&4.81(-11)&9.27(-11)&1.21(-10)&1.41(-10)\\
55,41 &5.66(-19)&5.21(-18)&2.77(-17)&2.75(-16)&6.23(-15)&9.66(-14)&3.66(-13)&8.15(-13)\\
55,43 &3.12(-13)&9.90(-13)&2.46(-12)&7.70(-12)&2.76(-11)&7.04(-11)&1.05(-10)&1.31(-10)\\
55,45 &2.42(-17)&1.86(-16)&8.49(-16)&6.49(-15)&9.69(-14)&1.00(-12)&3.11(-12)&6.20(-12)\\
55,47 &3.32(-16)&2.03(-15)&7.60(-15)&4.29(-14)&4.04(-13)&2.75(-12)&6.95(-12)&1.23(-11)\\
55,49 &3.26(-14)&8.47(-14)&1.63(-13)&3.69(-13)&8.94(-13)&2.21(-12)&4.81(-12)&8.61(-12)\\
55,51 &2.17(-14)&8.85(-14)&2.60(-13)&1.00(-12)&4.85(-12)&1.58(-11)&2.57(-11)&3.34(-11)\\
55,53 &3.12(-13)&9.90(-13)&2.46(-12)&7.70(-12)&2.76(-11)&7.04(-11)&1.05(-10)&1.31(-10)\\
57,41 &6.70(-20)&8.04(-19)&5.19(-18)&6.21(-17)&1.95(-15)&4.12(-14)&1.81(-13)&4.40(-13)\\
57,43 &4.14(-19)&4.62(-18)&2.80(-17)&3.12(-16)&8.54(-15)&1.55(-13)&6.06(-13)&1.36(-12)\\
57,45 &3.13(-18)&3.11(-17)&1.73(-16)&1.63(-15)&3.44(-14)&4.79(-13)&1.64(-12)&3.38(-12)\\
57,47 &4.32(-17)&3.48(-16)&1.63(-15)&1.21(-14)&1.73(-13)&1.68(-12)&4.93(-12)&9.35(-12)\\
57,49 &2.18(-15)&1.20(-14)&4.13(-14)&2.00(-13)&1.48(-12)&7.91(-12)&1.74(-11)&2.77(-11)\\
57,411&4.40(-14)&1.37(-13)&3.02(-13)&7.59(-13)&1.78(-12)&3.70(-12)&7.19(-12)&1.22(-11)\\
57,51 &3.06(-16)&2.16(-15)&9.32(-15)&5.57(-14)&5.07(-13)&2.91(-12)&6.05(-12)&8.98(-12)\\
57,53 &2.74(-15)&1.68(-14)&6.63(-14)&3.40(-13)&2.36(-12)&1.07(-11)&2.03(-11)&2.89(-11)\\
57,55 &5.70(-14)&2.50(-13)&8.00(-13)&3.15(-12)&1.45(-11)&4.61(-11)&7.56(-11)&9.97(-11)\\
\hline
\end{tabular}
\end{center}

\clearpage

\begin{table}[h]
\caption{Same as Table 2 but for the $v=6$ levle of  para-H$_2$.}
\end{table}
\begin{center}
\begin{tabular}{ccccccccc}\\
& \multicolumn{8}{c}{$T(K)$} \\ \hline\hline
${vj,v'j'}$&100&200&300&500&1000&2000&3000&4000\\ 
\hline
60,50 &3.91(-17)&2.58(-16)&9.12(-16)&4.70(-15)&4.39(-14)&3.92(-13)&1.20(-12)&2.36(-12)\\
60,52 &9.56(-17)&5.25(-16)&1.54(-15)&6.05(-15)&4.77(-14)&4.69(-13)&1.56(-12)&3.27(-12)\\
60,54 &7.24(-17)&4.31(-16)&1.61(-15)&1.01(-14)&1.06(-13)&7.79(-13)&2.21(-12)&4.29(-12)\\
60,56 &8.01(-17)&6.49(-16)&2.51(-15)&1.41(-14)&1.52(-13)&1.28(-12)&3.51(-12)&6.43(-12)\\
60,58 &1.06(-17)&9.97(-17)&3.72(-16)&3.62(-15)&8.20(-14)&1.22(-12)&4.05(-12)&7.86(-12)\\
62,50 &8.18(-18)&6.30(-17)&2.69(-16)&1.81(-15)&1.99(-14)&1.51(-13)&4.21(-13)&7.95(-13)\\
62,52 &5.18(-17)&3.84(-16)&1.57(-15)&9.81(-15)&1.03(-13)&8.42(-13)&2.46(-12)&4.76(-12)\\
62,54 &1.92(-16)&1.24(-15)&4.41(-15)&2.28(-14)&1.84(-13)&1.18(-12)&3.13(-12)&5.81(-12)\\
62,56 &6.88(-16)&3.38(-15)&9.41(-15)&3.55(-14)&2.49(-13)&1.71(-12)&4.43(-12)&7.89(-12)\\
62,58 &2.78(-15)&4.22(-15)&5.52(-15)&1.35(-14)&1.34(-13)&1.52(-12)&4.76(-12)&9.07(-12)\\
62,60 &4.62(-12)&9.70(-12)&1.52(-11)&2.58(-11)&4.67(-11)&7.02(-11)&8.32(-11)&9.15(-11)\\
64,50 &3.16(-18)&2.58(-17)&1.21(-16)&1.01(-15)&1.54(-14)&1.33(-13)&3.52(-13)&6.23(-13)\\
64,52 &2.48(-17)&1.92(-16)&8.51(-16)&6.55(-15)&8.78(-14)&7.09(-13)&1.87(-12)&3.34(-12)\\
64,54 &1.11(-16)&7.90(-16)&3.20(-15)&2.10(-14)&2.32(-13)&1.71(-12)&4.50(-12)&8.12(-12)\\
64,56 &6.66(-16)&4.00(-15)&1.38(-14)&6.94(-14)&5.40(-13)&3.13(-12)&7.42(-12)&1.26(-11)\\
64,58 &1.75(-14)&5.94(-14)&1.27(-13)&3.15(-13)&9.72(-13)&3.67(-12)&8.42(-12)&1.42(-11)\\
64,60 &8.78(-14)&3.00(-13)&6.61(-13)&1.77(-12)&5.70(-12)&1.31(-11)&1.82(-11)&2.15(-11)\\
64,62 &1.39(-12)&4.08(-12)&7.96(-12)&1.81(-11)&4.76(-11)&9.73(-11)&1.31(-10)&1.53(-10)\\
66,50 &7.42(-19)&7.00(-18)&3.71(-17)&3.85(-16)&8.71(-15)&1.04(-13)&2.96(-13)&5.26(-13)\\
66,52 &6.07(-18)&5.50(-17)&2.80(-16)&2.69(-15)&5.38(-14)&5.76(-13)&1.59(-12)&2.80(-12)\\
66,54 &2.98(-17)&2.50(-16)&1.17(-15)&9.73(-15)&1.55(-13)&1.37(-12)&3.56(-12)&6.12(-12)\\
66,56 &1.97(-16)&1.44(-15)&5.94(-15)&3.99(-14)&4.68(-13)&3.36(-12)&8.21(-12)&1.38(-11)\\
66,58 &2.82(-15)&1.60(-14)&5.18(-14)&2.42(-13)&1.73(-12)&8.66(-12)&1.83(-11)&2.86(-11)\\
66,510&2.94(-13)&6.00(-13)&9.23(-13)&1.55(-12)&2.85(-12)&6.08(-12)&1.16(-11)&1.83(-11)\\
66,60 &1.98(-15)&1.01(-14)&3.02(-14)&1.25(-13)&7.44(-13)&2.79(-12)&4.66(-12)&6.08(-12)\\
66,62 &2.23(-14)&1.05(-13)&2.86(-13)&1.03(-12)&5.25(-12)&1.77(-11)&2.86(-11)&3.68(-11)\\
66,64 &3.06(-13)&1.15(-12)&2.60(-12)&7.11(-12)&2.56(-11)&6.81(-11)&1.01(-10)&1.25(-10)\\
\hline
\end{tabular}
\end{center}

\clearpage

\begin{table}[h]
\caption{Same as Table 10 except for ortho-H$_2$.}
\end{table}
\begin{center}
\begin{tabular}{ccccccccc}\\
& \multicolumn{8}{c}{$T(K)$} \\ \hline\hline
${vj,v'j'}$&100&200&300&500&1000&2000&3000&4000\\ 
\hline
61,51 &2.13(-17)&1.66(-16)&7.25(-16)&5.02(-15)&6.40(-14)&6.64(-13)&2.21(-12)&4.57(-12)\\
61,53 &8.24(-17)&5.33(-16)&1.95(-15)&1.03(-14)&8.87(-14)&7.30(-13)&2.32(-12)&4.79(-12)\\
61,55 &1.22(-16)&5.99(-16)&1.92(-15)&1.03(-14)&1.23(-13)&1.13(-12)&3.25(-12)&6.16(-12)\\
61,57 &2.61(-16)&6.80(-16)&1.86(-15)&9.35(-15)&9.55(-14)&1.04(-12)&3.39(-12)&6.68(-12)\\
63,51 &9.64(-18)&7.36(-17)&3.26(-16)&2.43(-15)&3.49(-14)&3.55(-13)&1.11(-12)&2.21(-12)\\
63,53 &4.74(-17)&3.42(-16)&1.43(-15)&9.69(-15)&1.20(-13)&1.10(-12)&3.38(-12)&6.71(-12)\\
63,55 &2.32(-16)&1.47(-15)&5.38(-15)&2.95(-14)&2.68(-13)&1.84(-12)&4.84(-12)&8.84(-12)\\
63,57 &2.51(-15)&1.03(-14)&2.55(-14)&7.71(-14)&3.31(-13)&1.88(-12)&5.10(-12)&9.35(-12)\\
63,59 &9.22(-18)&1.47(-16)&8.79(-16)&7.31(-15)&7.79(-14)&8.44(-13)&3.06(-12)&6.47(-12)\\
63,61 &2.54(-12)&6.12(-12)&1.14(-11)&2.46(-11)&5.79(-11)&1.04(-10)&1.31(-10)&1.49(-10)\\
65,51 &2.30(-18)&2.07(-17)&1.02(-16)&9.05(-16)&1.72(-14)&2.19(-13)&7.27(-13)&1.47(-12)\\
65,53 &1.21(-17)&1.01(-16)&4.83(-16)&3.95(-15)&6.42(-14)&6.82(-13)&2.08(-12)&4.04(-12)\\
65,55 &7.34(-17)&5.52(-16)&2.38(-15)&1.67(-14)&2.15(-13)&1.89(-12)&5.34(-12)&9.96(-12)\\
65,57 &8.69(-16)&5.24(-15)&1.84(-14)&9.48(-14)&7.83(-13)&4.66(-12)&1.09(-11)&1.82(-11)\\
65,59 &7.70(-14)&1.97(-13)&3.61(-13)&7.44(-13)&1.63(-12)&3.86(-12)&7.95(-12)&1.33(-11)\\
65,61 &4.24(-14)&1.81(-13)&4.90(-13)&1.68(-12)&7.10(-12)&2.03(-11)&3.07(-11)&3.82(-11)\\
65,63 &4.98(-13)&1.68(-12)&3.91(-12)&1.10(-11)&3.54(-11)&8.29(-11)&1.18(-10)&1.43(-10)\\
67,51 &3.29(-19)&3.73(-18)&2.16(-17)&2.33(-16)&6.03(-15)&1.02(-13)&3.87(-13)&8.41(-13)\\
67,53 &1.58(-18)&1.63(-17)&1.04(-16)&1.08(-15)&2.44(-14)&3.50(-13)&1.18(-12)&2.41(-12)\\
67,55 &1.15(-17)&1.08(-16)&5.59(-16)&4.82(-15)&8.56(-14)&9.75(-13)&2.93(-12)&5.57(-12)\\
67,57 &1.40(-16)&1.09(-15)&4.62(-15)&3.08(-14)&3.74(-13)&3.08(-12)&8.17(-12)&1.45(-11)\\
67,59 &5.74(-15)&3.10(-14)&9.76(-14)&4.25(-13)&2.75(-12)&1.28(-11)&2.59(-11)&3.92(-11)\\
67,511&4.76(-14)&2.19(-13)&4.93(-13)&1.17(-12)&2.68(-12)&5.86(-12)&1.12(-11)&1.81(-11)\\
67,61 &8.54(-16)&6.05(-15)&2.27(-14)&1.19(-13)&9.11(-13)&4.19(-12)&7.78(-12)&1.08(-11)\\
67,63 &6.50(-15)&4.07(-14)&1.38(-13)&6.19(-13)&3.73(-12)&1.44(-11)&2.51(-11)&3.39(-11)\\
67,65 &1.06(-13)&4.94(-13)&1.39(-12)&4.74(-12)&1.96(-11)&5.63(-11)&8.72(-11)&1.11(-10)\\
\hline
\end{tabular}
\end{center}
\clearpage

\begin{figure}
\begin{center}
\epsfxsize=6in \epsfysize=7.2in \epsfbox{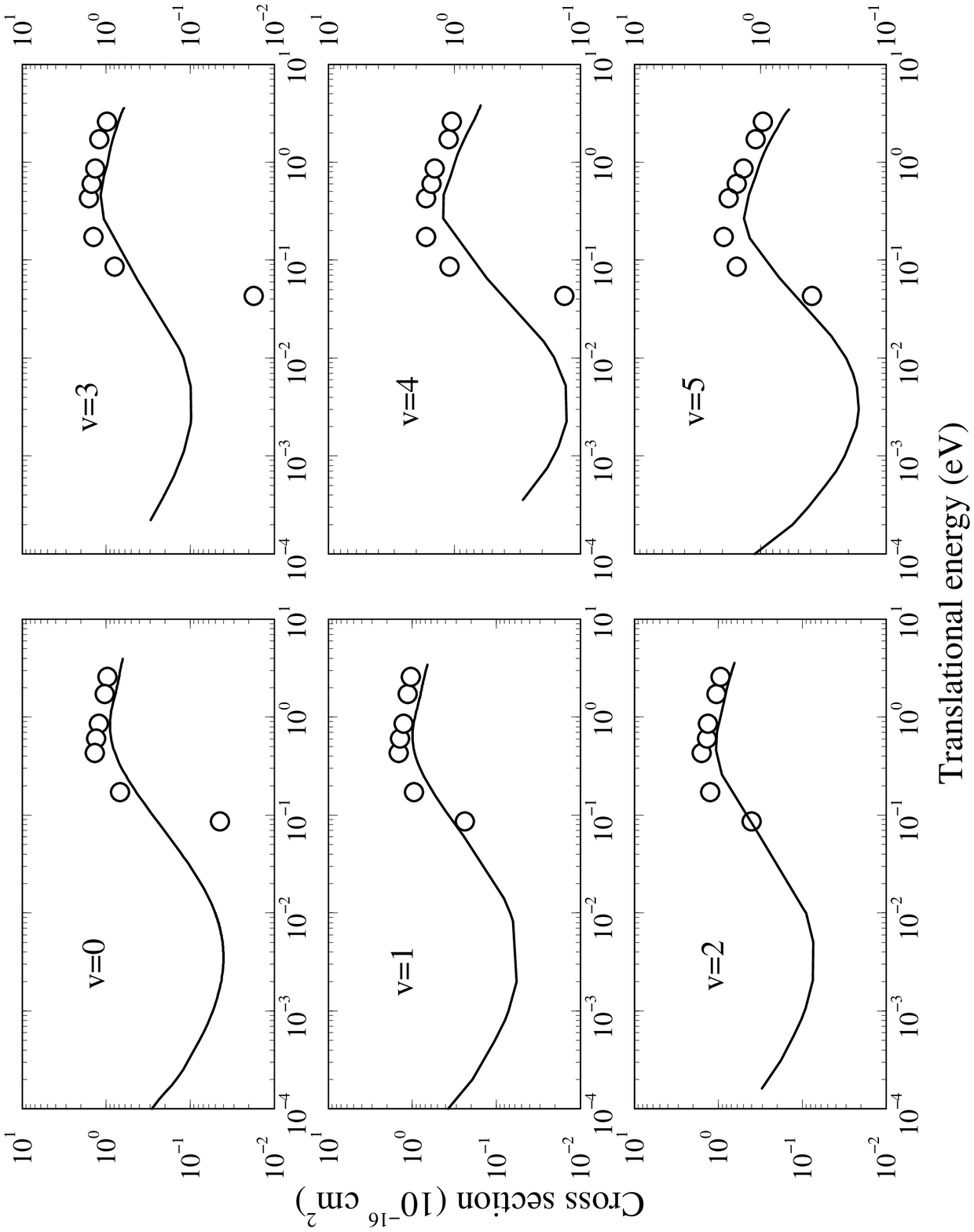}
\end{center}
\caption{Comparison of quantum mechanical (solid line) and 
quasiclassical cross sections (open circles) for
the $\Delta j=-2$ rotational transition from the $j=2$ initial 
rotational level of H$_2$ in 
vibrational levels $v=0$ to 5. The error bar for the classical mechanical
results is largest for the lowest energy where it is of the size of 
the symbol.} 
\end{figure}

\begin{figure}
\begin{center}
\epsfxsize=6in \epsfysize=7.2in \epsfbox{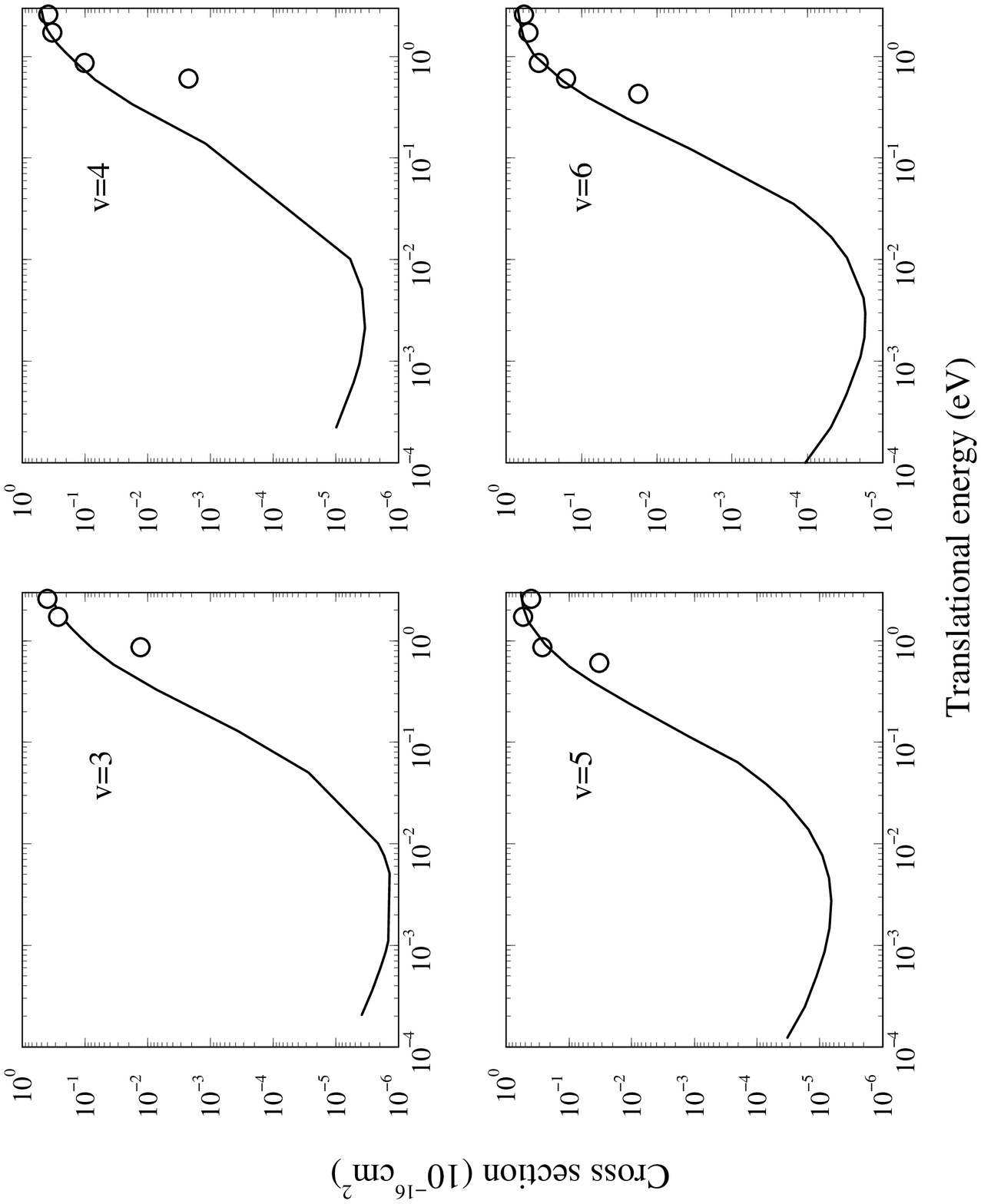}
\end{center}
\caption{Comparison of quantum mechanical (solid line) 
and quasiclassical cross sections (open circles) for
the $\Delta j=+2,\,\,\Delta v=-1$ rovibrational transition from the $j=5$ initial 
rotational level of H$_2$ in 
vibrational levels $v=3$ to 6.
The error bar for the classical mechanical
results is largest for the lowest energy where it is of the size of 
the symbol.} 
\end{figure}
\end{document}